\documentclass{iaus}
\usepackage{graphicx}
\usepackage{natbib}

  \checkfont{eurm10}
  \iffontfound
    \IfFileExists{upmath.sty}
      {\typeout{^^JFound AMS Euler Roman fonts on the system,
                   using the 'upmath' package.^^J}%
       \usepackage{upmath}}
      {\typeout{^^JFound AMS Euler Roman fonts on the system, but you
                   dont seem to have the}%
       \typeout{'upmath' package installed. iaus.cls can take advantage
                 of these fonts,^^Jif you use 'upmath' package.^^J}%
      }
  \else
  \fi

  \checkfont{msam10}
  \iffontfound
    \IfFileExists{amssymb.sty}
      {\typeout{^^JFound AMS Symbol fonts on the system, using the
                'amssymb' package.^^J}%
       \usepackage{amssymb}%

      }{}
  \fi

  \IfFileExists{amsbsy.sty}
    {\typeout{^^JFound the 'amsbsy' package on the system, using it.^^J}%
     \usepackage{amsbsy}}
    {\providecommand\boldsymbol[1]{\mbox{\boldmath $##1$}}}

\newsavebox{\astrutbox}
\sbox{\astrutbox}{\rule[-5pt]{0pt}{20pt}}

\newcommand\etal{\mbox{\textit{et al.}}}

\newcommand{\be}{\begin{equation}}
\newcommand{\ee}{\end{equation}}

\newcommand{\rhogas}{\rho_{\rm g}}

\newcommand{\BM}{$\beta$-model}
\newcommand{\bm}{\BM\ }

\newcommand{\Rv}{R_{200}}

\newcommand{\g}{{\sc GADGET}}
\newcommand{\lcdm }{$\Lambda $CDM }
\title[Outskirts of Galaxy Clusters: intense life in the suburbs]
      {The structure of the ICM from  High Resolution SPH simulations}

\author[G. Yepes \etal ]{G. Yepes$^1$, Y. Ascasibar$^2$, R. Sevilla$^1$, 
            S. Gottl\"ober$^3$  and V. M\"uller$^3$ }
\affiliation{$^1$Grupo de Astrof\'{\i}sica, Universidad Aut\'onoma de Madrid,
 Madrid E-280049, Spain \\
 $^2$Theoretical Physics, University of Oxford,
 1 Keble Road, Oxford OX1 3NP, United Kingdom \\
 $^3$Astrophysikalisches Institut Potsdam,
 An der Sternwarte 16, Potsdam D-14482, Germany}
\pubyear{2004}
\volume{195}
\pagerange{1--8}
\date{?? and in revised form ??}
\jname{Outskirts of Galaxy Clusters: intense life in the suburbs}
\editors{A. Diaferio, ed.}
\begin{document}

\maketitle

   \begin{abstract}
     We present  results from a set of high ($512^3$ effective
     resolution), and ultra-high ($1024^3)$  SPH
     adiabatic cosmological simulations of cluster formation
     aimed at  studying  the internal  structure of the
     intracluster medium (ICM).
     We discuss the radial structure and scaling relations expected
     from purely gravitational collapse, and show that the choice of a
     particular halo model can have important consequences on the
     interpretation of observational data.
     The validity of the 
      approximations of hydrostatic equilibrium and a
     polytropic equation of state are checked against 
      results of our  simulations. 
     We also show  the first  results from an unprecedented  
     large-scale  simulation of 500 $h^{-1}$ Mpc and $2\times 512^3$ 
      gas and dark matter particles. This experiment will make possible
       a detailed  study of the large-scale distribution of clusters  as a
      function of their  X-ray properties.
     \end{abstract}


%

\firstsection

\section{Introduction}

Galaxy clusters are  a unique laboratory to test the hierarchical
paradigm of structure formation. They are the best probes  of the large
scale structure of the Universe and have often been used as a
diagnostic of the cosmological parameters.
The intrinsic non-linear nature of gravitational collapse and
 gas dynamics makes numerical simulations the
most useful tool to study in detail the process of cluster formation
and evolution.


Simple analytical models for the structure of the ICM can be derived
from  the  hypotheses of  hydrostatic equilibrium and 
  polytropic equation of state, $P\propto\rhogas^\gamma$. But real clusters might not be well described by  
these two hypothesis. For instance, kinetic energy makes a 
significant contribution to the
energy budget of merging systems, and therefore thermally-supported
hydrostatic equilibrium ceases to be a valid approximation.
Even in relaxed systems, this assumption is not very accurate in the
outermost parts, where gas motions become more important.
Departures from spherical symmetry can also play a role in the final
structure of the ICM \citep{LeeSuto03} and, last but not
least, there is no obvious physical reason for the gas to follow a
polytropic relation. From our numerical experiments, we showed (see 
 \citet{Ascasibar03}) that hydrostatic equilibrium 
is fulfilled  within $\sim20$ per cent accuracy by all 
simulated clusters, as long
as they are not heavily disturbed.
A polytropic equation of state seems to be a good approximation as
well, although its reliability near the centre is still a matter of
debate \citep[see e.g.][]{Rasia_03}. From our data we derive a
polytropic index of $\gamma\sim 1.18$. 


We compared four different analytical  halo models with  our
simulations. 
 The first two models assume that haloes are well described
by  \citet{NFW97}
and  \citet{Moore99}  fitting formulae, 
while the other two assume that the gas follows a \bm
\citep{CavaliereFusco76}.
We consider a 'canonical' version of the \BM, in which the gas is
isothermal ($\gamma=1$) and $\beta=2/3$, and a polytropic version with 
$\gamma=1.18$ and $\beta=1$. The same value of the polytropic index
has been used for the first two models as well.
Hereafter we will use the abbreviations NFW, MQGSL, BM and PBM to
refer  to these models. For a detailed description, the
reader is referred to \citet{Ascasibar03}.

\section{Numerical experiments}

We have carried out a series of high-resolution gasdynamical
simulations of cluster formation in a flat $\Lambda$CDM universe 
($\Omega_{\rm m}=0.3$; $\Omega_\Lambda =0.7$; $h=0.7$; $\sigma_8=0.9$; $\Omega_{\rm b}=0.02\ h^{-2}$). 
Simulations have been done  with the parallel \g ~code
 \citep{gadget01}, with a novel version of SPH in which the entropy
  is explicitly conserved \citep{Gadget02}. 
In a  cubic volume of 80 $h^{-1}$ Mpc on a side, an unconstrained
realization of the  power spectrum of density fluctuations
corresponding to the \lcdm model was generated for a
total of $1024^3$ Fourier modes. The  density field was then resampled
to a grid of $128^3$  particles, which were displaced from their
Lagrangian positions according to the Zeldovich approximation up to $z=49$.
Their evolution until the present epoch is traced by means of a pure
N-body simulation with $128^3$ dark matter particles. A sample of
clusters  selected from this preliminary low-resolution experiment were
 re-simulated with higher resolution by means of
the multiple mass technique \citep[see][for details]{Klypin01}.
 Mass resolution is then increased by using smaller masses 
in the Lagrangian volume depicted  by these particles, 
including the additional small-scale waves from the \lcdm power
spectrum  in the new initial conditions.
We use 3 levels of mass refinement, reaching an effective resolution
of $512^3$ CDM particles ($2.96\times 10^8\ h^{-1}$
M$_\odot$). Gas was   added in the highest resolved area
only. The total number of particles (dark+SPH) in this area is
greater than $1-2\times 10^6$ for all clusters.
The gravitational softening length was set to $\epsilon=2-5\ h^{-1}$
kpc, depending on number of particles within the virial radius 
\citep{Power03}. The minimum smoothing length for SPH was fixed 
to the same value as $\epsilon$.

In order to study effects of resolution in the determination of
X-ray  properties of  our clusters, we have resimulated
one of the objects with  8 times more mass resolution, reaching an
effective resolution of $1024^3$ particles (i.e. $m_{dark} \sim 3\times 10^7\
h^{-1} M_\odot$; $m_{sph}\sim 5\times 10^6 M_\odot$). The total number of
particles within the virial radius was  $11,106,465$.

The list of our simulated clusters extracted from the 80 h$^{-1}$ Mpc volume 
span a relative small range in  X-ray emission temperature (from 0.6 to
3 keV). In order to extend our numerical sample of clusters  to  wider
temperature (mass) range, we have simulated a considerable much bigger
volume  (500 h$^{-1}$ Mpc) in which a random realization of the \lcdm power
spectrum was generated with $2048^3$   particles. In this way, we will
have a similar resolution for our clusters than in the previous
experiments. We have resimulated the whole 500 h$^{-1}$ Mpc box with different
mass resolutions: $2\times 128^3$, $2\times 256^3$, and $2\times512^3$ dark and SPH
particles. We identified all halos in the lowest resolution run
($128^3$). Using the same technique as before, we  selected
clusters in this  run  and resimulated  them at full resolution
($2048^3$), using 5 different species of dark particles and SPH  with 
$m_{sph}\sim 1.6\times 10^8 h^{-1}  M_\odot$. The clusters  selected cover
a range of masses from $2\times 10^{14} - 2\times  10^{15}$  $M_\odot$. In this
regard, we could  extend the temperature range of our cluster sample  up
to 11 keV.  The total number of particles within virial radius in these
clusters is comparable to the number of particles of previous
simulations ($\sim 10^6$).

\begin{figure*}
  \centering \includegraphics[width=4.4cm]{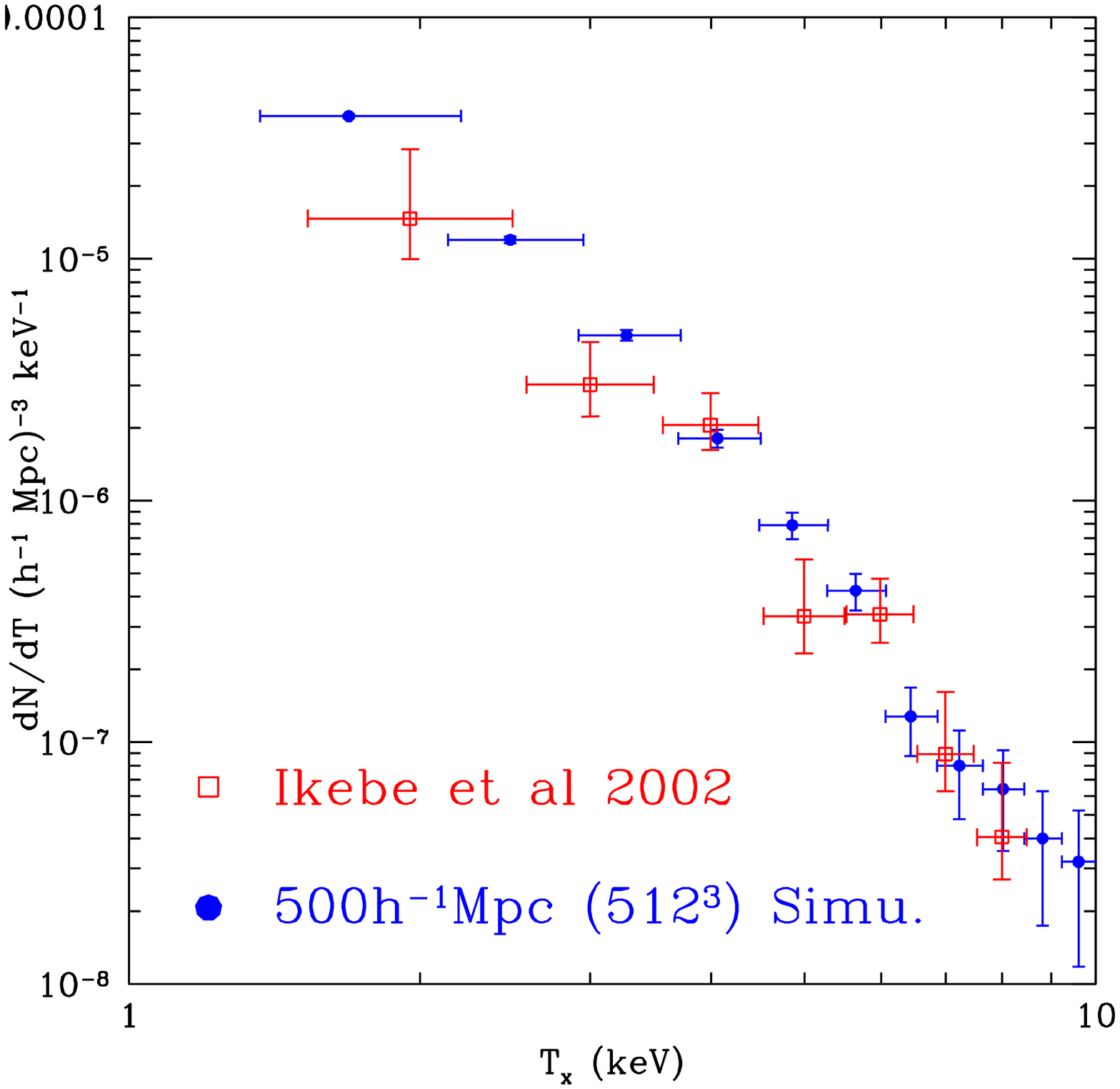}\includegraphics[width=4.4cm]{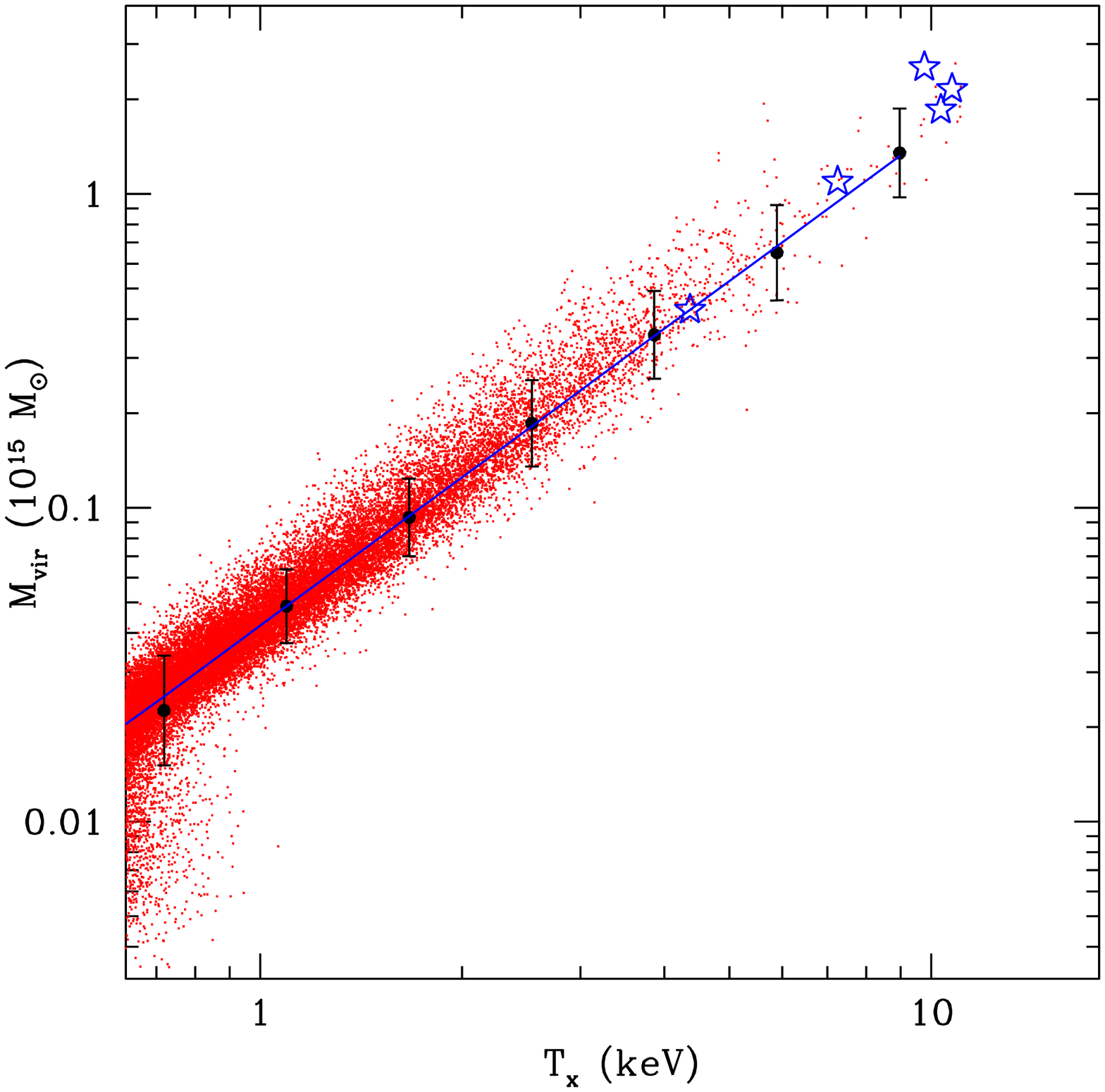}\includegraphics[width=4.4cm]{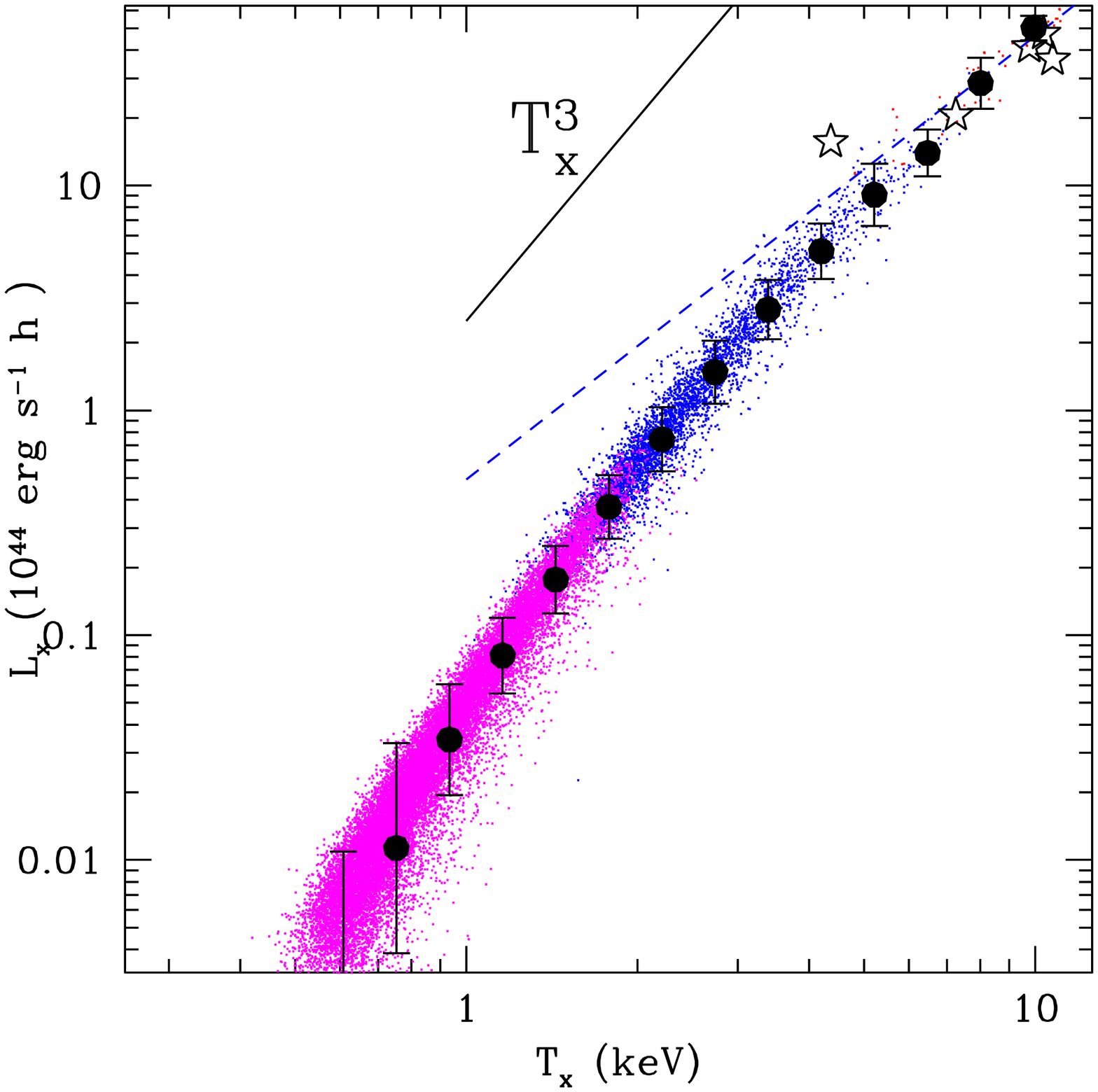}
  \caption{Results from $500^3$ simulated volume. Left, the X-ray
    Temperature function and its observational estimates \citep[]{ikebe02}. 
 Middle, the $M_{vir}-T_x$ relation. Right,
    the $L_X-T_X$ relation. Stars represent the clusters resimulated
    with high resolution. The dotted line is a fit to the hottest
    halos ( $> 6$ keV). The slope is $\sim 2$, as in the self-similar
    scaling behaviour}  
  \label{xtf}
\end{figure*}

We have recently finished the  run  with $2\times 512^3$ gas and dark
particles that, to our knowledge, is one of the largest  adiabatic SPH
simulations of large scale structure done so far. The mass resolution
is $m_{dark}\sim 6\times 10^{10} h^{-1} \, M_\odot$, which means that we can
resolve  from  galactic halos  (100+100 particles)  to the biggest galaxy
clusters ($4\times 10^6$ particles).  We identified a total of $4\times 10^5$
dark matter halos,  with 10 or more particles, in this
simulation.  Well resolved halos (7000 dark particles or more)
correspond to  clusters  with emission weighted temperatures  $> 3$
 keV.  If we go down to 1000 particles, the halos resolved have temperatures
$> 1$ keV, and  temperatures go down to $\sim 0.6$ keV for halos with 500
or more particles. Due to the large simulated volume, we have a
statistical significant  sample of clusters and groups. The total
number of objects with $T_X>0.6$ keV, excluding substructure, is $\sim
30,000$. A total of 126 hot clusters have been found with $T_X >5$ keV.
The X-ray temperature function  for clusters in this simulation is
shown in Figure  \ref{xtf}, together with recent observational 
estimates.  As can be seen, the number density of clusters for the
highest temperatures ($> 6$ keV)  found in simulations is compatible with
observations (see \citet[]{Borgani04} for  results  from 
non-adiabatic simulations).  

The resulting $M-T_X$ and $L_X-T_X$ relations are depicted  in
Figure \ref{xtf}, in which we also show 6 of the hotter  clusters
resimulated with high resolution. They show a convergence of results at
least for clusters with $T_X > 3 $ keV.  The fit to the $M-T_X$ relation
for clusters with $T_X > 1$ keV is
$(M_{vir}/M_0) =  (T_x/\mbox{keV})^\alpha $, with $\alpha=1.56\pm 0.05$ and
$M_0=4.4\times 10^{13} h^{-1}  M_\odot $.  The slope is compatible with
simulations which include non adiabatic effects, although the zero
point, $M_0$, is a factor of 2 higher. Note however that we are using virial
mass instead of  mass  for overdensity 500. This implies  that the M-T
relation is rather insensitive  to the energy transfers due to non
adiabatic processes associated to star-gas interactions. 

\begin{figure}
  \centering \includegraphics[width=7cm]{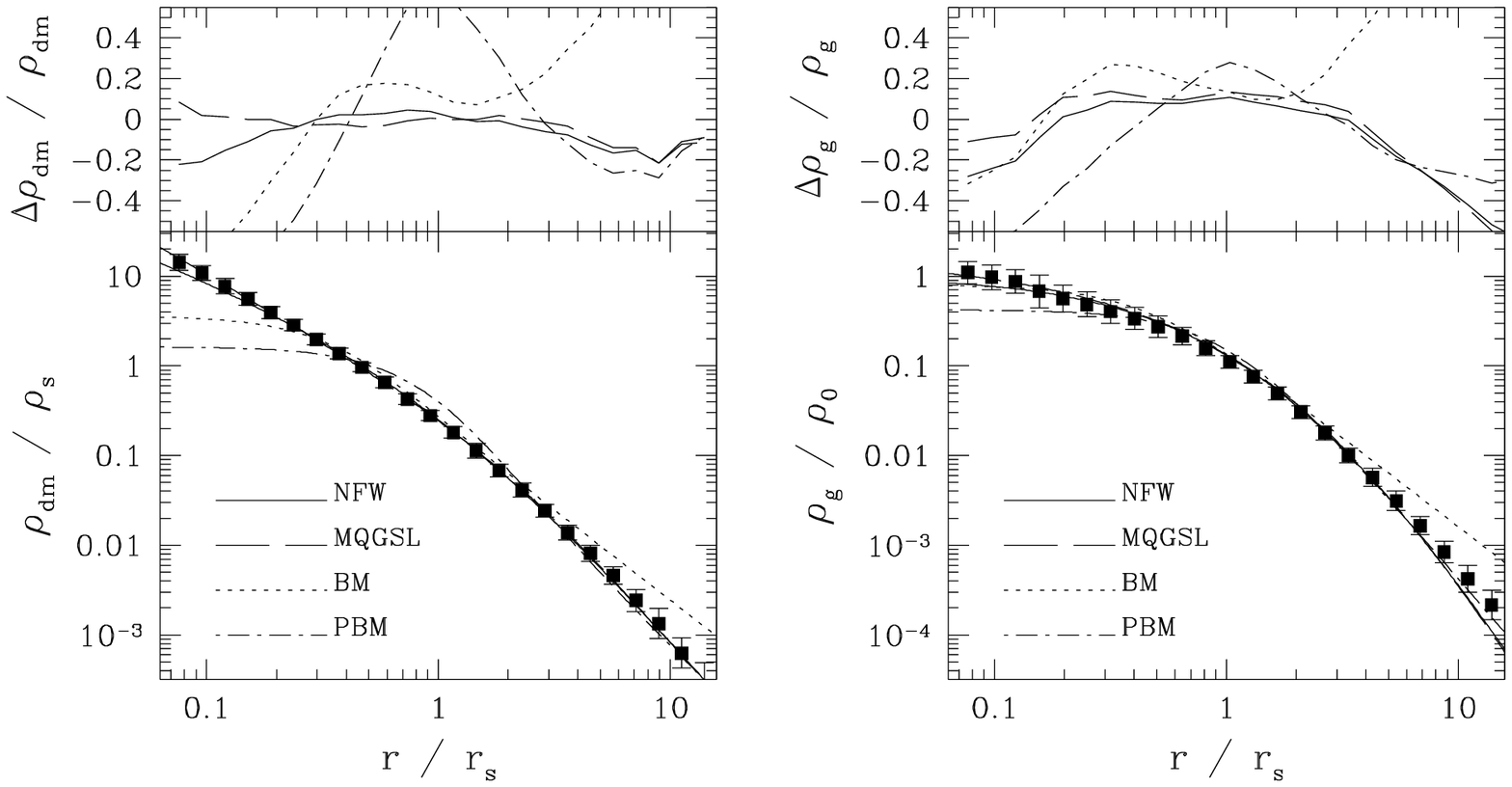}

\centering \includegraphics[width=3.5cm,clip]{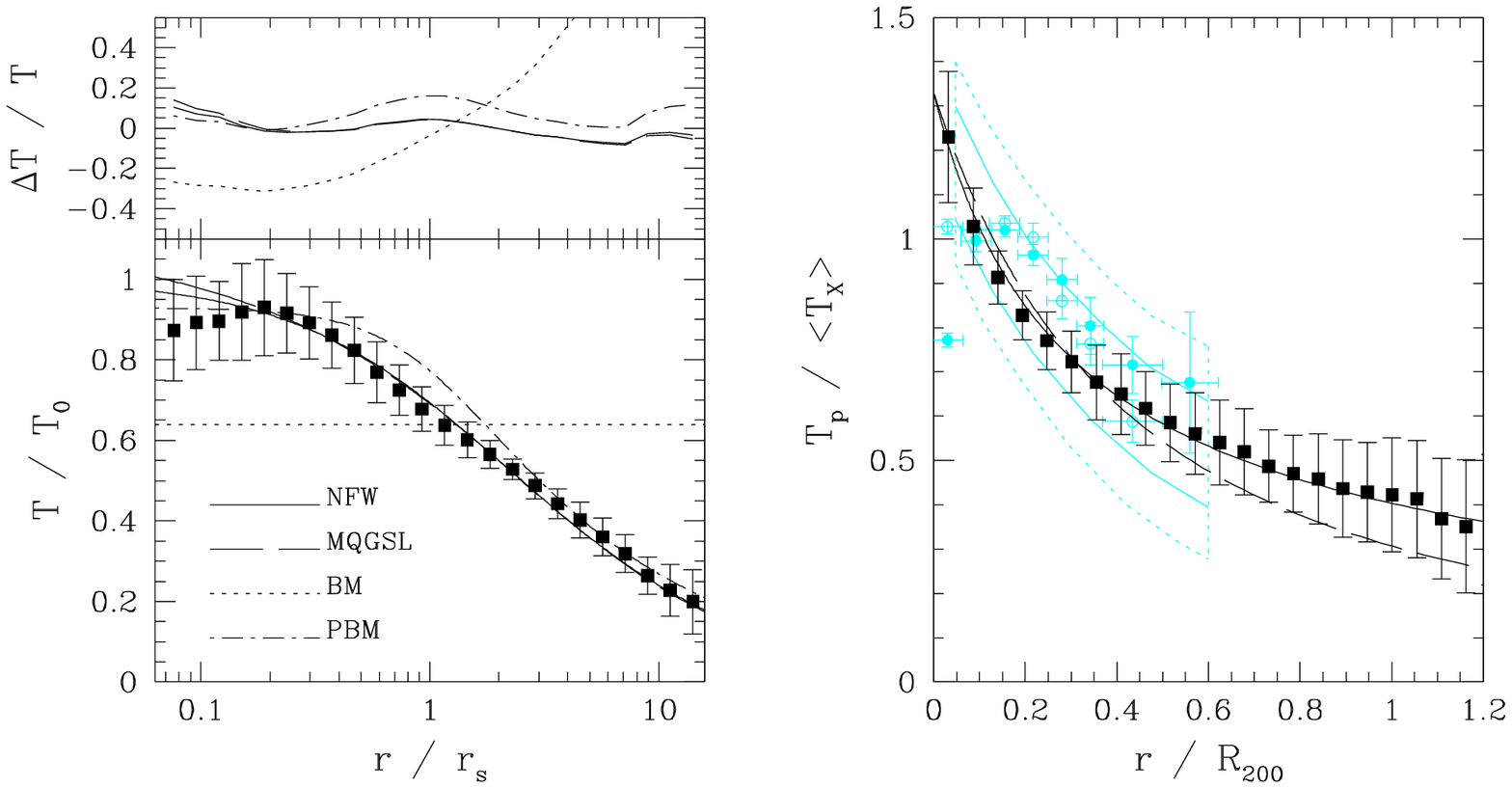} 
\hspace*{0.12cm}
\includegraphics[width=3.35cm]{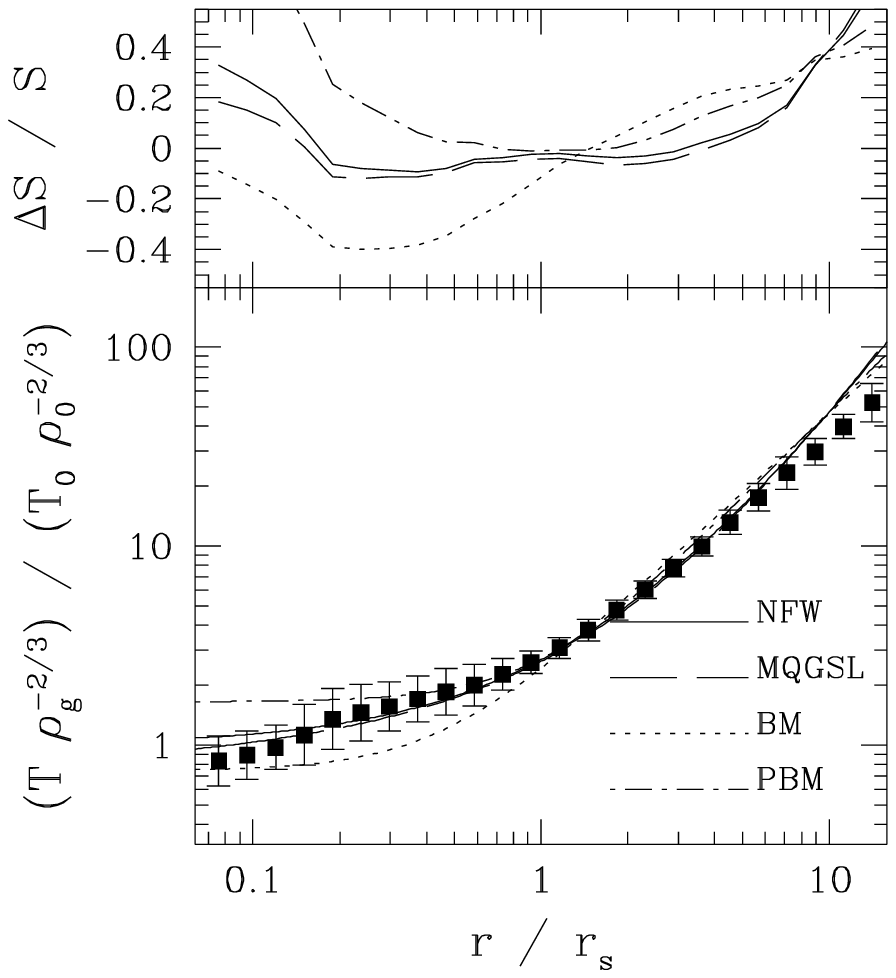} 
  \caption{Universal radial profiles: Upper panel  dark matter and
    gas  density. Lower panel:  projected temperature and gas entropy.
Black squares represent the numerical data, averaged over all clusters 
except major mergers. Error bars denote one-sigma scatter. Lines are
used  to plot the analytical models, and top panels quantify deviations from
 the simulated profile.  Dashed lines in the  projected
 emission-weighted temperature profile,  shows the
 'universal' profile proposed by \citet{Loken02}, 
while solid line represents the best fit to our data. 
Observations by \citet[circles]{GrandiMolendi02} and \citet[boxed region]{Markevitch98} are ploted for comparison.}
  \label{figDens}
\end{figure}

\section{Radial structure of the ICM}

We compare the universal  halo models described earlier, 
 with results from our simulations.
For NFW and MQGSL, we obtain the characteristic density and radius 
of each cluster by fitting the dark matter distribution. 
Gas density and temperature profiles are genuine predictions of these models.
For BM and PBM, we fit the gas distribution and predict the gas 
temperature and the dark matter density.
In Figure~\ref{figDens} we plot  the average dark matter and gas density
 profiles. Not surprisingly, both NFW and MQGSL provide good fits to
 the  dark matter density.
  These models are able to accurately predict the gas density, 
but they are too steep at large radii due to a systematic departure 
from hydrostatic equilibrium.

On the other hand, \BM s (BM and PBM) show a core in the dark matter 
density that is not seen in the  numerical data.
 Moreover, they do not give  an accurate description of the gas 
density profile. The inner regions are better described with low values
of $\beta$, while the outer parts require higher values for this parameter.



One of the most striking results from our simulations is that  they
favour  the hypothesis of a 'universal' temperature profile. 
As can be seen in Figure~\ref{figDens}, the ICM is not isothermal,
 but the temperature decreases by a factor of three or four from the
 centre  to the virial radius.  
%
The projected X-ray emission-weighted temperature profile is plotted 
on the right panel of Figure~\ref{figDens}. 
We compare our simulations with previous work by \citet{Loken02} based
 on Eulerian simulations. These authors propose a 'universal' form
$ T_{\rm p}(r)=T_0{(1+r/a_{\rm x})}^{-\delta}$
Our results are well described by
this relation.
Although real clusters seem to be consistent with a polytropic 
equation of state \citep{Markevitch98}, recent observations indicate
the presence of a large isothermal core \citep{GrandiMolendi02}. 
Apart from this feature, which is not observed in our objects, 
adiabatic simulations are in good agreement with observational 
data beyond $0.2\Rv$.




Entropy profiles are plotted on the lower panel of Figure~\ref{figDens}.
Contrary to the common view, neither the analytical models nor the
simulation  data yield a pure power-law profile, despite the fact that purely
adiabatic gasdynamics has been considered. As shown in
\citet{Ascasibar03},
 the standard implementation of the SPH algorithm can lead to misleading
results in the inner regions due to artificial entropy losses
 \citep[see e.g.][]{Gadget02}.
We find that the shape of the entropy profile does not depend on the
cluster  mass or temperature,
in agreement with recent observations \citep{Ponman03}.
NFW or MQGSL models provide a better estimate of the entropy
profile than the \BM, but the low gas densities
predicted at large radii yield a very steep slope at $r\sim\Rv$.


\section{Convergence of results}
\begin{figure*}
  \centering \includegraphics[width=6cm]{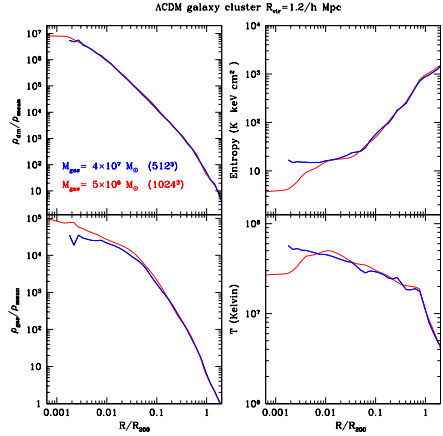}
  \caption{Radial profiles for a cluster run with  two different resolutions. }
  \label{profres}
\end{figure*}

To check for convergence of results in terms of numerical resolution,
 we compare, in Figure \ref{profres} 
  radial profiles for one of our clusters  from the ultra-high
  resolution simulation described in previous section.
As it can be appreciated, the structure of the ICM is quite similar, at
least from the virial radius down to 1\% of that. The features at the
most internal parts of the cluster are related with the different
dynamical stage of the cluster.
The positive results of this comparison is that the overall properties
of the halos are well described with the resolution adopted in our
simulations in which  500,000-1,000,000  SPH particles are used.

\begin{acknowledgements}
This work has been partially supported by the MCyT (Spain, AYA-0973),
 by the {\em Acciones Integradas
Hispano-Alemanas} HA2000-0026 and by DAAD (Germany). We thank the 
 Forschungszentrum J\"ulich for allowing us to use the IBM p690+
 supercomputer. 
\end{acknowledgements}

\bibliographystyle{aa}
\bibliography{DATABASE,PREPRINTS}

\begin{thebibliography}{16}
\expandafter\ifx\csname natexlab\endcsname\relax\def\natexlab#1{#1}\fi

\bibitem[{{Ascasibar} {et~al.}(2003){Ascasibar}, {Yepes}, {M{\" u}ller}, \&
  {Gottl{\" o}ber}}]{Ascasibar03}
{Ascasibar}, Y., {Yepes}, G., {M{\" u}ller}, V., \& {Gottl{\" o}ber}, S. 2003,
  \mnras, 346, 731

\bibitem[{{Borgani} {et~al.}(2004){Borgani}, {Murante}, {Springel}, {Diaferio},
  {Dolag}, {Moscardini}, {Tormen}, {Tornatore}, \& {Tozzi}}]{Borgani04}
{Borgani}, S., {Murante}, G., {Springel}, V., {et~al.} 2004, \mnras, 348, 1078

\bibitem[{{Cavaliere} \& {Fusco-Femiano}(1976)}]{CavaliereFusco76}
{Cavaliere}, A. \& {Fusco-Femiano}, R. 1976, \aap, 49, 137

\bibitem[{{De Grandi} \& {Molendi}(2002)}]{GrandiMolendi02}
{De Grandi}, S. \& {Molendi}, S. 2002, \apj, 567, 163

\bibitem[{{Ikebe} {et~al.}(2002){Ikebe}, {Reiprich}, {B{\" o}hringer},
  {Tanaka}, \& {Kitayama}}]{ikebe02}
{Ikebe}, Y., {Reiprich}, T.~H., {B{\" o}hringer}, H., {Tanaka}, Y., \&
  {Kitayama}, T. 2002, \aap, 383, 773

\bibitem[{{Klypin} {et~al.}(2001){Klypin}, {Kravtsov}, {Bullock}, \&
  {Primack}}]{Klypin01}
{Klypin}, A., {Kravtsov}, A.~V., {Bullock}, J.~S., \& {Primack}, J.~R. 2001,
  \apj, 554, 903

\bibitem[{{Lee} \& {Suto}(2003)}]{LeeSuto03}
{Lee}, J. \& {Suto}, Y. 2003, \apj, 585, 151

\bibitem[{{Loken} {et~al.}(2002){Loken}, {Norman}, {Nelson}, {Burns}, {Bryan},
  \& {Motl}}]{Loken02}
{Loken}, C., {Norman}, M.~L., {Nelson}, E., {et~al.} 2002, \apj, 579, 571

\bibitem[{{Markevitch} {et~al.}(1998){Markevitch}, {Forman}, {Sarazin}, \&
  {Vikhlinin}}]{Markevitch98}
{Markevitch}, M., {Forman}, W.~R., {Sarazin}, C.~L., \& {Vikhlinin}, A. 1998,
  \apj, 503, 77

\bibitem[{{Moore} {et~al.}(1999){Moore}, {Quinn}, {Governato}, {Stadel}, \&
  {Lake}}]{Moore99}
{Moore}, B., {Quinn}, T., {Governato}, F., {Stadel}, J., \& {Lake}, G. 1999,
  \mnras, 310, 1147

\bibitem[{{Navarro} {et~al.}(1997){Navarro}, {Frenk}, \& {White}}]{NFW97}
{Navarro}, J.~F., {Frenk}, C.~S., \& {White}, S.~D.~M. 1997, \apj, 490, 493

\bibitem[{{Ponman} {et~al.}(2003){Ponman}, {Sanderson}, \&
  {Finoguenov}}]{Ponman03}
{Ponman}, T.~J., {Sanderson}, A.~J.~R., \& {Finoguenov}, A. 2003, \mnras, 343,
  331

\bibitem[{{Power} {et~al.}(2003){Power}, {Navarro}, {Jenkins}, {Frenk},
  {White}, {Springel}, {Stadel}, \& {Quinn}}]{Power03}
{Power}, C., {Navarro}, J.~F., {Jenkins}, A., {et~al.} 2003, \mnras, 338, 14

\bibitem[{{Rasia} {et~al.}(2003){Rasia}, {Tormen}, \& {Moscardini}}]{Rasia_03}
{Rasia}, E., {Tormen}, G., \& {Moscardini}, L. 2003, {\tt astroph/0309405}


\bibitem[{{Springel} \& {Hernquist}(2002)}]{Gadget02}
{Springel}, V. \& {Hernquist}, L. 2002, \mnras, 333, 649

\bibitem[{{Springel} {et~al.}(2001){Springel}, {Yoshida}, \&
  {White}}]{gadget01}
{Springel}, V., {Yoshida}, N., \& {White}, S.~D.~M. 2001, New Astronomy, 6, 79

\end{thebibliography}

\end{document}